\documentclass[runningheads]{llncs}

\usepackage{url}
\usepackage{cite}
\usepackage{graphicx}
\usepackage{cite}
\usepackage{indentfirst}
\usepackage[table]{xcolor}
\usepackage{float}
\usepackage{amsmath,amssymb,amsfonts}
\usepackage{algorithmic}
\usepackage{graphicx}
\usepackage{textcomp}
\usepackage{xcolor}
\usepackage{graphicx}
\usepackage{algorithmic}
\usepackage[linesnumbered,ruled]{algorithm2e}
\usepackage{psfrag}

\usepackage{color, colortbl}
\definecolor{Gray}{gray}{0.9}
\definecolor{DarkGray}{gray}{0.4}
\def\BibTeX{{\rm B\kern-.05em{\sc i\kern-.025em b}\kern-.08em
    T\kern-.1667em\lower.7ex\hbox{E}\kern-.125emX}}

\begin{document}
\title{Resilient Consensus via Weight Learning and Its Application in Fault-Tolerant Clock Synchronization \thanks{This work
was supported in part by the National Natural Science Foundation of
China (NSFC) under grant numbers 61803340, 61751210, 61673344 and
61701444.}}
%
%
\author{Jian Hou$^{1,2}$ \and
Zhiyong Chen$^{2}$ \and ZhiyunLin$^{3}$ \thanks{Corresponding
author.} \and Mengfan Xiang$^{1}$}

\authorrunning{Hou et al.}
%
\institute{$^1$School of Information Science, Zhejiang Sci-Tech University \\
Hangzhou 310018, China \\
$^2$School of Electrical Engineering and Computing, The University
of Newcastle, Callaghan, NSW 2308, Australia\\
$^3$School of Automation, Hangzhou Dianzi University, Hangzhou
310018, China \\
\email{changeleap@163.com, zhiyong.chen@newcastle.edu.au, linz@hdu.edu.cn}\\
}
\maketitle
\begin{abstract}
This paper addresses the distributed consensus problem in the
presence of faulty nodes.   A novel weight learning algorithm is
introduced such that neither network connectivity nor a sequence of
history records is required to achieve resilient consensus.   The
critical idea is to dynamically update the interaction weights among
neighbors learnt from their credibility measurement. Basically, we
define a reward function that is inversely proportional to the
distance to its neighbor, and then adjust the credibility based on
the reward derived at the present step and the previous credibility.
In such a way, the interaction weights are updated at every step,
which integrates the historic information and degrades the
influences from faulty nodes. Both fixed and stochastic topologies
are considered in this paper. Furthermore, we apply this novel
approach in clock synchronization problem. By updating the logical
clock skew and offset via the corresponding weight learning
algorithms, respectively, the logical clock synchronization is
eventually achieved regardless of faulty nodes. Simulations are
provided to illustrate the effectiveness of the strategy.

\end{abstract}
\section{Introduction}
Multi-agent consensus is a fundamental problem in distributed
systems, and has been studied for decades in the fields of computer
science, control, communication and many others. The objective is to
achieve global agreement with local collaborative interactions. In
practice, faulty agents with non-cooperative behaviors are
inevitable to interfere system coordination, such as Internet
ghostwriters to influence public opinions, enemy aircrafts to
disturb unmanned aerial vehicle formation control, malfunctioning
clocks to break clock synchronization, and so on \cite{FTR19}. How
to effectively identify all the faulty nodes and eliminate their
negative effects so as to achieve resilient consensus is a
challenging problem.

A typical method is to remove the extreme neighbor nodes at state
updating. This method is available only if the network connectivity
is not less than $2F +1$, where $F$ is the maximal number of faulty
nodes \cite{DFCV11,CCIU12}. In \cite{CIN11}, the authors presented a
Mean-Subsequence-Reduced (MSR) algorithm that combines ideas of
distributed computing and control consensus protocols to solve the
asymptotic consensus under the $F$-total malicious model in complete
networks. This MSR algorithm was further generalized to the
Weighted-Mean-Subsequence-Reduced (W-MSR) algorithm to solve both
malicious and Byzantine threat models \cite{LCR12}. Following the
idea, an asynchronous delayed network case was considered
\cite{RCSO17}. In \cite{IAB12}, the authors provided tight
conditions for resilient consensus using the MSR algorithm. In
addition, the authors of \cite{RAC13} proposed a novel topological
property where network robustness replaces the network connectivity
metrics. This kind of methods by removing the extreme neighbor nodes
are constrained to the network connectivity or network robustness
and hardly realized in practice.

Another way to solve the consensus problem with faulty nodes is to
evaluate the trustworthiness on each neighbor node. The
trustworthiness is typically determined by a weight measuring the
influence of the neighbor nodes. In \cite{TMF12}, the authors
presented an algorithm named RoboTrust, to calculate trustworthiness
of agents using observations and statistical inferences from various
historical perspectives, and thus all the agents finally converge to
the value agreed by the most trustworthy agents. The trustworthiness
established by local evidence in \cite{TMF12} was extended to the
second-hand evidence in \cite{UTI14}. The trustworthiness based
consensus strategy with faulty nodes avoids the network connectivity
or network robustness, yet requires to store and analyze a mass of
historical information.

To overcome the disadvantage of requiring network connectivity or
network robustness, or storage and analysis of historical
information, this paper presents a Weight Learning Algorithm (WLA)
borrowing the idea from Reinforcement Learning (RL)
\cite{SRA19,MAML14}. In a multi-agent consensus problem, adjacent
weights between two neighbor nodes indicate the strength of
influence. Therefore, the goal is to isolate the faulty nodes by
reducing the adjacent weights to zero in a local consensus protocol.
To achieve it, we first construct a reward function by the relative
distance of neighbor nodes. Next, each credibility is updated by its
previous value and the reward at the present step. Finally, all the
adjacent weights are determined by the credibility via
normalization. In this way, the adjacent weights from faulty nodes
become smaller compared to the ones from normal nodes, and tend to
zero.

We consider three different classes of nodes in this paper,
represented by normal nodes, persistent faulty nodes
\cite{FTCS17,RAC13}, and intermittent faulty nodes. Persistent
faulty nodes may take arbitrary uncertain states at each time due to
faults or external attacks. Intermittent faulty nodes, which is a
mixed behavior of normal nodes and that of persistent faulty nodes
with a certain probability,  inject hostile influence to the network
intermittently by not being easily detected. Furthermore, we apply
this novel approach in a clock synchronization problem.

Clock synchronization is a common and fundamental problem in
wireless sensor networks (WSNs), for various applications such as
scheduling, information fusion, and so on. The substantial results
in analyzing this problem include Reference Broadcast
Synchronization \cite{FGN02}, Timing-sync Protocol for Sensor
Networks \cite{TSP03}, Lightweight Time Synchronization, Flooding
Time Synchronization Protocol \cite{SFT13}, and Consensus Based
Synchronization Protocol \cite{AT11,TSI13,NCS13,ACS17,FTCS17}. Among
these, the consensus-based approach is  popular in recent years due
to its scalability, robustness, distributed manner, and simple
implementability.

The consensus-based protocols can be classified into two categories,
i.e., Average Time Synchronization (ATS) \cite{AT11} and Maximum
Time Synchronization (MTS) \cite{TSI13}. In ATS, each node utilizes
the average value of neighbor nodes and its own to achieve
consensus, and in MTS, this value becomes the maximum of the
neighbors and the node itself. In \cite{NCS13}, the authors present
the second-order consensus strategy to solve the clock
synchronization problem under measurement noises and time-varying
clock drifts. \cite{ACS17} proposes a modified MTS algorithm to
speed up convergence under bounded noise. \cite{FTCS17} borrows the
idea of MSR algorithm to remove the outliers in the clock data
received from the neighbors to deal with the unreliable channels in
WSNs.

In this paper, we consider the clock synchronization problem with
bounded system noise. As some nodes in the network may be hacked and
transmit arbitrary signals (occasionally) by not following the given
protocol, we construct logical clock skew and offset, and update
their values via WLA. As a result, the faulty nodes are gradually
isolated and all the normal nodes achieve resilient logical clock
synchronization.

The contributions of this paper are as follows:
\begin{enumerate}
\item[(1)] Most references assume various network conditions, such as $F$-total
fault model (up to $F$ faulty nodes), $F$-local fault model (up to
$F$ faulty incoming neighbors for each normal node), or network
connectivity (no less than $2F+1$).  The proposed algorithm only
requires a rooted communication topology among normal nodes.

\item[(2)] Some references without assuming the aforementioned network conditions require a
mass of historical information to analyze, that is hard to realize
in large scale networks. The proposed algorithm is simple and
efficient by integrating all the historical information into one
variable.

\item[(3)] Two different misbehaved models are considered for both fixed
and stochastic topologies in this paper.

\item[(4)] The proposed approach is applied in a clock
synchronization problem with hacked or malfunctioning nodes and
system noise,  and hence achieves resilient logical clock
synchronization for all normal nodes.
\end{enumerate}

The rest of this paper is organized as follows.
Section~\ref{sec:background} introduces necessary preliminaries and
system models. The detailed WLAs for both fixed communication
topology and stochastic communication topology are given in
Section~\ref{sec:main1}. Section~\ref{sec:appl} discusses the clock
synchronization problem and the corresponding solution by our
approach. Numerical simulations and conclusions are presented in
Sections~\ref{sec:simu} and \ref{sec:con}, respectively.

\section{Preliminaries and Problem Formulation}\label{sec:background}

In this paper, we consider the consensus problem in a point-to-point
message-passing network, which is thus modeled as a directed graph
(digraph). Given a digraph $G = (V,E,A)$ where $V=\{1,2,\cdots,n\}$
is the node set, $E \subseteq V \times V$ is the edge set, and $A =
\{ [a_{ij}] \geq 0 \}$ represents the weighted adjacency matrix. An
edge $(j,i) \in E$ exists if and only if there is information flow
from node $j$ to node $i$, i.e., $a_{ij} > 0$. It is assumed
$a_{ii}=0$ with no self-loop. The neighbor set of node $i$ is
presented by $N_i=\{ j \; |\; (j,i) \in E \}$. A path from node $j$
to $i$ is a sequence of distinct nodes $i_0, i_1, \cdots, i_m$,
where $i_0=j$ and $i_m=i$ with $(i_l,i_{l+1}) \in E, 0 \leq l \leq
m-1$. We say a digraph rooted if there exists a node $i \in V$,
called root, such that there is a path from node $i$ to any other
node. We use the terms node and agent interchangeably. The notations
$a_{ij}(k)$ and $N_i(k)$ are used when they vary with time $k$.


We consider $n$ agents with a discrete-time system model
\begin{align}\label{basicmodeldis}
x_i(k+1)=x_i(k)+u_i(k), \; i \in V
\end{align}
where $x_i(k) \in \mathbb{R}$ and $u_i(k) \in \mathbb{R}$ represent
the state and control input of node $i$, respectively. The initial
state of the $n$ nodes, i.e.,
$\mathbf{x}(0)=[x_1(0),\cdots,x_n(0)]^T$, is arbitrarily specified.

In a traditional consensus problem, all nodes are assumed equally
trustworthy to cooperate with each other to achieve state consensus
\cite{HouZheng18,PWC17}, through proper design of $u_i(k)$. In this
paper, we consider three different classes of nodes, represented by
the set of normal nodes $V^{\rm n}$, the set of persistent faulty
nodes $V^{\rm p}$, and the set of intermittent  faulty nodes $V^{\rm
i}$, with
\begin{align*}
V= V^{\rm n}\cup V^{\rm p}\cup V^{\rm i}.
\end{align*}
The faulty nodes include passive ones caused by system failure and
active adversarial ones that deliberately inject hostile influence
to the network. The control actions for the three sets of nodes are
described as follows.

 \begin{enumerate}

\item[(i)] \textbf{Normal Node}:
The following consensus algorithm is executed in a normal node,
\begin{align}\label{basicmodeldis1}
u_i(k)= \!\!\!\sum_{j \in N_i(k)}\!\! a_{ij}(k)(x_j(k) - x_i(k))+
\omega_i(k),\; i \in V^{\rm n},
\end{align}
where $a_{ij}(k)$ is the adjacent weight from node $j$ to node $i$
satisfying $\sum_{j \in N_i(k)} a_{ij}(k) < 1$, and $\omega_i(k)
\in \mathbb{R}$ is a bounded noise ($|\omega_i(k) |< \omega$)
introduced by transmission channel and environment.


\item[(i)] \textbf{Persistent Faulty Node (PFN)}:
A PFN conducts its input with a random value at every time as
follows
 \begin{align}
u_i(k)= {\rm Random}, \; i \in V^{\rm p},
\end{align}
where $\rm Random$ is a random variable that has a specified
probability density function $f_{\rm Random}$.

\item[(ii)] \textbf{Intermittent Faulty Node (IFN)}:
An IFN mixes the behavior of a normal node and that of a PFN with a
certain probability, that is,
\begin{align*}
u_{i}(k)=\left\{
\begin{array}{ll}
\sum\limits_{j \in N_i(k)} a_{ij}(k)(x_j(k) - x_i(k))+ \omega_i(k), \\
&\hspace{-4cm}\mbox{with probability } p\\
{\rm Random}, &\hspace{-1cm}  \mbox{else}
\end{array} \right. , \\  i \in V^{\rm i}.
\end{align*}
It especially describes the behavior of a node that acts normally
most time not to be detected but intermittently disturbs the network
using random actions.
 \end{enumerate}

\begin{remark}
It is worth mentioning that the assumption $a_{ij}(k) \in \{ 0 \}
\cup [c,1]$ with $c$ a positive constant is widely used in the
traditional consensus settings; see, e.g., \cite{CMC05}. That is,
the nonzero weights must be sufficiently away from zero with a lower
bound. However, in the present setting, this assumption is not
applied as a mechanism is designed to deliberately tune the weights
associated with faulty nodes, ideally to zero, to mitigate their
influence on the desired network behavior.  \end{remark}

\begin{remark}
The general actions of faulty nodes like PFN and/or IFN as most
frequently referred is taken into consideration in this paper. The
intelligent faulty nodes with antagonistic behaviors that can add
clever corruptions to avoid detection by the given updating scheme
is not covered in this paper, and would be our future research
interest.
\end{remark}

With the appearance of faulty nodes (PFN and/or IFN), the whole
network does not achieve consensus in general. This paper aims to
design a distributed algorithm for updating network weights among
nodes such that the faulty nodes can be isolated from the normal
nodes with the weights from the former to the latter can be tuned to
be close to zero, that is
\begin{align}
\limsup_{k\rightarrow\infty} \max_{i \in V^{\rm n}, j \in V^{\rm p}
\cup V^{\rm i}} |a_{ij}(k)| < \epsilon \label{isolation}
\end{align}
for a sufficiently small $\epsilon$. As a result, we expect that the
sub-network of normal nodes can still achieve consensus in the
following sense
\begin{align}
\limsup_{k\rightarrow\infty} \max_{i, j \in V^{\rm n}}
|x_{i}(k)-x_j(k)| < \varepsilon \label{consensus}
\end{align}
for a sufficiently small $\varepsilon$. Throughout the paper, it is
assumed that   the sub-network of normal nodes is rooted in a fixed
topology (or with probability one in a stochastic topology).

The achievement of \eqref{consensus}  for the system
\eqref{basicmodeldis} in the presence of PFN and/or IFN is called
resilient consensus in this paper.  So, the main {\it objective} is
to propose a distributed algorithm for updating network weights to
achieve resilient consensus.


\section{Weight Learning Algorithms}\label{sec:main1}

A distributed algorithm for updating network weights in the
aforementioned objective is called a Weight Learning Algorithm (WLA)
in this section. The idea  used in the WLA architecture is borrowed
from RL using the so-called reward. In particular, for each normal
node $i$ and its neighbor $j$, we use a three-level learning
strategy:  an immediate reward $r_{ij}$ describing the performance
at the current instant, a credibility $Q_{ij}$  integrating the
historical  trustworthiness up to present, and an updating rule for
the corresponding adjacent weight $a_{ij}$. Thus, we expect to
adjust and reduce the adjacent weights from faulty nodes to normal
nodes through trial-and-error interactions, and hence mitigate the
influence of faulty nodes for the final objective of resilient
consensus. In the sequel, we will study two cases: a fixed topology
case and a stochastic topology case.

\subsection{Fixed Topology Case}
We first consider the WLA in a network with a fixed communication
topology, abbreviated to WLA-F. In the algorithm, a critical concept
is a {\it reward}  denoted by $r_{ij}(k)$ for all normal nodes $i\in
V^{\rm n}$ and with their neighbors $j\in N_i$. In particular, a
normal node recognizes its own state $x_i$ as the true value and
evaluates the reward by the relative state of a neighbor node,
represented by the following reward function
\begin{align*}
r_{ij}(k)= f(|x_j(k)-x_i(k)+\omega_{ij}(k)|, k), \; j\in N_i ,\;i\in
V^{\rm n}
\end{align*}
where $\omega_{ij}$ is the unknown noise on the transmission channel
from $j$ to $i$ with $|\omega_{ij}| < \omega$. In general, we can
select a reward function $f$ as an inverse proportional function,
e.g.,
\begin{align*}
f(|x_j(k)-x_i(k)+\omega_{ij}(k)|,k)
=e^{-|x_j(k)-x_i(k)+\omega_{ij}(k)| \theta(k)}
\end{align*}
for an appropriately designed parameter $\theta(k) > 0$. With such a
selection, we restrict the value of the reward function between $0$
and $1$.  When the states $x_i$ and $x_j$ are sufficiently
different, the induced reward is close to $0$; when they agree, the
reward moves towards $1$.

Based on the reward, we define a {\it credibility} $Q_{ij}$  between
two nodes,
\begin{align*}
Q_{ij}(k)=Q_{ij}(k-1) r_{ij}(k),\; Q_{ij}(0)=1, \; j\in N_i ,\;i\in
V^{\rm n}.
\end{align*}
It is  initialized as $1$ and then recursively updated according to
the associated reward $r_{ij}$. A credibility $Q_{ij}$ essentially
integrates all the historical information from node $j$ to node $i$.
It contains a mechanism that the credibility reduces more
significantly if the reward is closer to $0$ due to the major
difference between the node states.

Now, it is ready to find a strategy for weight updating and hence a
complete WLA-F. By normalization of the credibility of all neighbors
for a normal node, we define the weight as
\begin{align*}
a_{ij}(k)= \frac{Q_{ij}(k)}{\sum_{j  \in N_i}
Q_{ij}(k)}(1-\frac{1}{|N_i|}), \;  j \in N_i ,\;i\in V^{\rm n},
\end{align*} where $|N_i|$ is the cardinality of $N_i$. The design
obviously satisfies $\sum_{j \in N_i} a_{ij}(k) < 1$.

It is noted that $a_{ij}$ is not applicable for PFNs and it does not
have to be updated for IFNs. For the complement of notation, the
following simple rule is applied
\begin{align} \label{aijfaulty}
a_{ij}(k)= a_{ij}(0), \;  j \in N_i ,\;i\in V^{\rm p} \cup V^{\rm
i}, \end{align} with arbitrarily initialized $a_{ij}(0)$ satisfying
$\sum_{j \in N_i} a_{ij}(0) < 1$. Lastly, from the definition of
$N_i$, it is trivially known that
\begin{align}  \label{aij0}
a_{ij}(k)=0,\; j \notin N_i ,\;i\in V.
\end{align}

From the above, the WLA-F is summarized in Algorithm~1 using
pseudocode. Also, the closed-loop system takes the following form
 \begin{align}
x_{i}(k+1)= & x_i(k)  +    \sum_{j  \in N_i} a_{ij}(k)(x_j(k) -
x_i(k)) \nonumber\\ & +\omega_{i}(k),\; i \in V^{\rm n}, \nonumber \\
x_{i}(k+1)&=x_i(k)+ {\rm Random}, \; i \in V^{\rm p}, \label{close}
\end{align}
and the mixture for $i \in V^{\rm i}$ with the normal node behavior
of probability $p$.

\begin{algorithm}
\caption{WLA-F for a normal node $i\in V^{\rm n}$} \label{Alg1}
\footnotesize \label{alg:WLAFC}
\begin{algorithmic}[1]


\STATE Initialize $Q_{ij}(0)=1$, $j\in N_i$  \\

\FOR{$k=1; \  k++$}  \FOR{$j=1; \ j<n+1; \ j++$}

\IF {$j\in N_i$}

\STATE   $r_{ij}(k)= f(|x_j(k)-x_i(k)+\omega_{ij}(k)|)$;  \\
$Q_{ij}(k)=Q_{ij}(k-1) r_{ij}(k)$;\\
$a_{ij}(k)= \frac{Q_{ij}(k)}{\sum_{j  \in N_i}
Q_{ij}(k)}(1-\frac{1}{|N_i|})$;

 \ELSE
\STATE $a_{ij}(k)= 0$; \ENDIF \ENDFOR

\STATE $x_{i}(k+1)=x_i(k)+ \sum\limits_{j  \in N_i} a_{ij}(k)(x_j(k)
- x_i(k))+\omega_{i}(k)$.

 \ENDFOR
\end{algorithmic}

\end{algorithm}

\subsection{Stochastic Topology Case}
In this subsection, we consider the WLA in a stochastic
communication topology (WLA-S) where all the communications are
stochastic.  To tailor the algorithm for this scenario, we need to
modify the reward, the credibility, and the weight update rule
accordingly.

As the topology is stochastic, a communication edge may occur
between a normal node and any other node. Therefore, we must
maintain the rewards $r_{ij}(k)$ for a normal node $i\in V^{\rm n}$
and all other nodes $j\in V$. Noting $N_i(k)$ is now time-varying,
one has
\begin{align*}
r_{ij}(k)= \left\{\begin{array}{ll}
f(|x_j(k)-x_i(k)+\omega_{ij}(k)|,k), & j\in N_i(k) \\
r_{ij}(k-1), & j\notin N_i(k) \end{array} \right. ,\\  i\in V^{\rm
n}
\end{align*}
with the initial setting
 \begin{align*}
r_{ij}(0)=1,\; j\in V ,\;i\in V^{\rm n}.
\end{align*}
In other words, at each time instant, the reward $r_{ij}(k)$ updates
in a normal way if node $j$ is a neighbor of node $i$ but keeps the
historical value $r_{ij}(k-1)$ otherwise.

With the modified definition of reward, the credibility $Q_{ij}$ is
modified accordingly,
\begin{align*}
Q_{ij}(k)= Q_{ij}(k-1) r_{ij}(k),\; i\in V^{\rm n},
\end{align*}
with the initial values
\begin{align*}
Q_{ij}(0)=1, \; j\in V ,\;i\in V^{\rm n}.
\end{align*}

The weight is thus updated following the rule
\begin{align*}
a_{ij}(k)=  \frac{Q_{ij}(k)}{\sum_{j \in N_i(k)} Q_{ij}(k)} \gamma
,\; j \in N_i(k), \; i\in V^{\rm n}
\end{align*}
where $\gamma \in (0,1)$ is an appropriately selected parameter. It
contains a mechanism that the updating of node $i$ relies more on
neighbor nodes if the value $\gamma$ is close to $1$, and results
probably in a faster convergence rate, which yet may produce
unstable performance. It is assured that $\sum_{j \in N_i(k)}
a_{ij}(k) < 1$ always holds. The update rules \eqref{aijfaulty} and
\eqref{aij0} are slightly revised as follows
\begin{align*}
a_{ij}(k)&= a_{ij}(0), \;  j \in N_i(k) ,\;i\in V^{\rm p} \cup V^{\rm i},  \\
a_{ij}(k)&=0,\; j \notin N_i(k) ,\;i\in V.
\end{align*}
with arbitrarily initialized $a_{ij}(0)$ satisfying $\sum_{j \in V}
a_{ij}(0) < 1$.

Finally,  the WLA-S is summarized in Algorithm~2 using pseudocode.
Also, the closed-loop system takes the same form as that in the
fixed topology case.

\begin{remark}
It is worth mentioning that $\gamma=1-\frac{1}{|N_i|}$ in fixed
topology case. We abandon the usage of $\gamma=1-\frac{1}{|N_i(k)|}$
in stochastic topology case to avoid that $|N_i(k)|=0$ or $1$.
\end{remark}

\begin{algorithm}
\caption{WLA-S for a normal node $i\in V^{\rm n}$} \label{Alg2}
\footnotesize \label{alg:WLASC}
\begin{algorithmic}[1]


\STATE Initialize $Q_{ij}(0)=1$ and $r_{ij}(0)=1$, $j\in V$;

\FOR{$k=1; \  k++$}  \FOR{$j=1; \ j<n+1; \ j++$}

\IF {$j\in N_i(k)$}

\STATE   $r_{ij}(k)= f(|x_j(k)-x_i(k)+\omega_{ij}(k)|)$;  \\
$Q_{ij}(k)=Q_{ij}(k-1) r_{ij}(k)$;\\
$a_{ij}(k)= \frac{Q_{ij}(k)}{\sum_{j  \in N_i(k)} Q_{ij}(k)}\gamma$;

 \ELSE
\STATE
$r_{ij}(k) =r_{ij}(k-1)$;\\
$Q_{ij}(k) =Q_{ij}(k-1) r_{ij}(k)$;\\
 $a_{ij}(k)= 0$;
\ENDIF \ENDFOR

\STATE $x_{i}(k+1)=x_i(k)+ \sum\limits_{j  \in N_i} a_{ij}(k)(x_j(k)
- x_i(k))+\omega_{i}(k)$.

 \ENDFOR
\end{algorithmic}

\end{algorithm}

\section{Application to Clock Synchronization}\label{sec:appl}

In this section, we introduce the clock synchronization problem and
use it to demonstrate the applicability of the proposed WLA. In
particular, it shows that WLA is able to achieve clock
synchronization in WSNs when some nodes are faulty or behave
abnormally under attack.

\subsection{Clock Model}

In this paper, we consider a group of $n$ linear clock models
\cite{TSI13,ACS17}, each representing  a hardware clock whose
reading at time $t$ is
\begin{align}
\tau_i^*(t) = \alpha_i^* t + \beta_i^*.
\end{align}
Here, $\alpha_i^*$ represents the clock skew that determines the
clock speed, and $\beta_i^*$ denotes the clock offset. Since the
clock skews may be slightly different from each other owing to
imperfect crystal oscillators, ambient temperature, battery voltage,
or oscillator aging \cite{TSI13}, we assume that $\alpha_i^*$ is
normalized around 1. It should be noted that the true values of both
parameters $\alpha_i^*$ and $\beta_i^*$ can not be obtained as the
absolute time $t$ is not accessible to the nodes. Therefore, two new
parameters $\alpha_i(t)$ and $\beta_i(t)$ are introduced to produce
a logical clock value
\begin{align*}
\tau_i(t) = \alpha_i(t) \tau_i^*(t) + \beta_i(t)= \alpha_i(t)
\alpha_i^* t +  \alpha_i(t) \beta_i^* + \beta_i(t).
\end{align*}
We call
\begin{align*}
x'_i(t) = \alpha_i(t) \alpha_i^*,\; x''_i(t) = \alpha_i(t) \beta_i^*
+ \beta_i(t)
\end{align*}
the logical clock skew and offset, respectively.

\begin{remark}
In practice, the clock skew $\alpha_i^*$ and the clock offset
$\beta_i^*$ vary slowly with time growing. It is expected that the
induced errors can be compensated if the proposed synchronization
algorithm converges fast enough.
\end{remark}

In the system, all the nodes exchange their current information with
neighbors periodically with a fixed period $T$. This means that the
real period $T_i$ for each node $i$ is $T_i=T/\alpha_i^*$. The
information exchanged includes the index $i$, original time
$\tau_i^*(t) $, logical time $\tau_i(t) $, and two parameters
$\alpha_i(t)$ and $\beta_i(t)$. We update the parameters at time
instant $t_k$ with $k \in \mathbb{Z}^+$ and use time index $k$ with
a slight abuse of notation. It is assumed that the communication
topology for normal nodes is rooted for each time interval $T/\min_i
\alpha_i^*$.

Now,  the objective is to design distributed control laws for
$\alpha_i(k)$ and $\beta_i(k)$ using the communication networks
represented by two weighted adjacency matrices $A' =\{ [a'_{ij}]
\geq 0 \}$ and $A'' =\{ [a''_{ij}] \geq 0 \}$, respectively, to
achieve logical clock synchronization in the sense of
\begin{align*}
\lim_{k\rightarrow\infty} \max_{i, j \in V} |x'_{i}(k)-x'_j(k)| &= 0,  \\
\lim_{k\rightarrow\infty} \max_{i, j \in V} |x''_{i}(k)-x''_j(k)| &=
0.
\end{align*}
Moreover, when PFNs and IFNs are present, the objective is  to
verify the proposed WLA for achieving resilient consensus in the
sense of \eqref{isolation} and \eqref{consensus} with both $a = a'$,
$x=x'$ and $a= a''$, $x=x''$.

\subsection{Logical Clock Skew Consensus}

The following distributed consensus algorithm is designed for
$\alpha_i(k)$ in a normal node $i$,
\begin{align}\label{skewmodel}
\alpha_i(k+1)=& \alpha_i(k) +  \sum_{j \in N_i(k)}
a'_{ij}(k)(\eta_{ij} \alpha_j(k) - \alpha_i(k)  \nonumber\\ & +
w'_{ij} (k)) ,\; i\in V^{\rm n}
\end{align}
where
\begin{align*}
\eta_{ij}=\frac{\tau_j^*(k_1)-\tau_j^*(k_2)}{\tau_i^*(k_1)-\tau_i^*(k_2)}
\end{align*}
is an estimate of ratio $\alpha_j^*/\alpha_i^*$, $k_1 \neq k_2$. The
behavior of a PFN is trivially governed by
\begin{align}\label{skewmodelPFN}
\alpha_i(k+1)= \alpha_i(k) +  {\rm Random},\; i \in V^{\rm p}
\end{align}
and that of an IFN  $i \in V^{\rm i}$ is the mixture with the normal
node behavior of probability $p$.


The following calculation
\begin{align*}
&\alpha_i(k+1)\alpha_i^* \\ =& \alpha_i(k)\alpha_i^* + \sum_{j \in
N_i(k)} a'_{ij}(k)[\eta_{ij} \alpha_j(k)\alpha_i^* -
\alpha_i(k)\alpha_i^* \\ &  + \alpha_j(k)\alpha_j^* -
\alpha_j(k)\alpha_j^* +w'_{ij}(k)\alpha_j^*]\\ =&
\alpha_i(k)\alpha_i^* + \sum_{j \in
N_i(k)}\!\!  a'_{ij}(k)[ \alpha_j(k)\alpha_j^* - \alpha_i(k)\alpha_i^* \\
& + (\eta_{ij}\frac{\alpha_i^*}{\alpha_j^*} - 1)
\alpha_j(k)\alpha_j^* +w'_{ij}(k)\alpha_j^* ]\\ =&
\alpha_i(k)\alpha_i^* +  \sum_{j \in N_i(k)}  a'_{ij}(k)[
\alpha_j(k)\alpha_j^* - \alpha_i(k)\alpha_i^* \\&
+w'_{ij}(k)\alpha_j^*]
\end{align*}
implies the following model for the logical clock skew,
\begin{align} \label{skewmodelnormal2}
x'_i(k+1)=& x'_i(k)+ \sum_{j \in N_i(k)} a'_{ij}(k)(x'_j(k)-x'_i(k))
+ \nonumber\\ & +w'_i(k) ,\; i\in V^{\rm n}.
\end{align}
Also, direct calculation on \eqref{skewmodelPFN} gives
\begin{align}\label{skewmodelPFN2}
x'_i(k+1)= x'_i(k) +  \alpha_i^* {\rm Random},\; i \in V^{\rm p}.
\end{align}
It is noted that $w'_{i}(k)= \sum_{j \in N_i(k)}
w'_{ij}(k)\alpha_j^*$ represents new noise and $\alpha_i^* {\rm
Random}$ a new random variable. The closed-loop systems
\eqref{skewmodelnormal2} and \eqref{skewmodelPFN2} take the form of
\eqref{close} with $a = a'$, $x=x'$. Therefore,  the proposed WLA in
the previous section applies for achieving resilient consensus with
the reward function, with $\bar w'_{ij}(k) = w'_{ij}(k)\alpha_j^*$,
\begin{align*}
&f(|x'_j(k)-x'_i(k)+\bar w'_{ij}(k) |, k)\\
=& f(\alpha_i^* | \eta_{ij}\alpha_j(k)  -\alpha_i(k)+w'_{ij}(k)  |, k)\\
= & f'( | \eta_{ij}\alpha_j(k)  -\alpha_i(k)+w'_{ij}(k)  |, k),
\end{align*}
which essentially relies on transmission of the information
$\eta_{ij} \alpha_j(k) - \alpha_i(k)+w'_{ij}(k)$ between neighbor
nodes subject to noise and the unknown $\alpha_i^*$ is absorbed by
the design of the function $f$.

\subsection{Logical Clock Offset Consensus}

Next, the distributed control law for $\beta_i(k)$ in a normal node
is given by
\begin{align}\label{offsetmodel}
\beta_i(k+1)= \beta_i(k) +  \sum_{j \in N_i(k)}
a''_{ij}(k)[\tau_j(k) \nonumber\\ - \alpha_i(k) \tau_i^*(k)
-\beta_i(k) +w''_{ij}],\; i\in V^{\rm n}.
\end{align}
The behavior of a PFN is trivially governed by
\begin{align}\label{offsetmodelPFN}
\beta_i(k+1)= \beta_i(k) +  {\rm Random},\; i \in V^{\rm p}.
\end{align}

Using the following fact,
\begin{align*}
\tau_j(k)   - \alpha_i(k) \tau_i^*(k)  -\beta_i(k)  \\
=x_j'(k) k + x_j''(k)  - x_i'(k) k - x_i''(k),
\end{align*}
one has
\begin{align*}
x''(k+1)= & \alpha_i(k+1)\beta_i^*+ \beta_i(k+1)\\
= &\alpha_i(k)\beta_i^* + \beta_i(k)
+[\alpha_i(k+1) -\alpha_i(k) ]\beta_i^* \\
 & +  \sum_{j \in N_i(k)}
a''_{ij}(k)[x_j'(k) k + x_j''(k)  - x_i'(k) k \\& - x_i''(k)  +w''_{ij}]\\
=& x''(k) +  \sum_{j \in N_i(k)}
a''_{ij}(k)[ x_j''(k) - x_i''(k) ]  \\
& +[x'_i(k+1) -x'_i(k) ]\beta_i^*/\alpha_i^* \\
& +  \sum_{j \in N_i(k)} a''_{ij}(k)[(x_j'(k) - x_i(k)) k
+w''_{ij}].
\end{align*}
As a result,
\begin{align} \label{offsetmodelnormal2}
x''(k+1) = & x''(k) +   \sum_{j \in N_i(k)} a''_{ij}(k)[ x_j''(k) -
x_i''(k) ]   \nonumber \\ & +w''_i ,\; i\in V^{\rm n}.
\end{align}
Also, direct calculation on \eqref{skewmodelPFN} and
\eqref{offsetmodelPFN} gives
\begin{align} \label{offsetmodelPFN2}
x''(k+1)=x''(k)+   \beta_i^* {\rm Random}+   {\rm Random},\; i \in
V^{\rm p}.
\end{align}
It is noted that, in \eqref{offsetmodelnormal2},
\begin{align*}
w''_i =& [x'_i(k+1) -x'_i(k) ]\beta_i^*/\alpha_i^* \\
&+  \sum_{j \in N_i(k)} a''_{ij}(k)[(x_j'(k) - x'_i(k)) k +w''_{ij}]
\end{align*}  represents the noise depending on the convergence of clock skew consensus;
and, in \eqref{offsetmodelPFN2},   $\beta_i^* {\rm Random}+   {\rm
Random}$ is a new random variable. The closed-loop systems
\eqref{offsetmodelnormal2} and \eqref{offsetmodelPFN2} take the form
of \eqref{close} with $a = a''$, $x=x''$. Therefore,  the proposed
WLA also applies for achieving resilient consensus with the reward
function, with $\bar w''_{ij}(k) = x_j'(k) k  - x_i'(k) k
+w''_{ij}(k)$,
\begin{align*}
&f (| x_j''(k)- x_i''(k)  +\bar w''_{ij} (k) |, k)\\
=& f( |\tau_j(k)   - \alpha_i(k) \tau_i^*(k)  -\beta_i(k)
+w''_{ij}(k) |, k ),
\end{align*}
which essentially relies on the transmission of the information
$\tau_j(k)   - \alpha_i(k) \tau_i^*(k)  -\beta_i(k) +w''_{ij}(k)$
between neighbor nodes subject to noise.

\section{Numerical Validation}\label{sec:simu}
In this section, we propose several numerical experiments to verify
the algorithms. Throughout the experiments, we let the reward
function $f(s,k)=e^{-s \theta(k)}$ with $\theta(k)=10^{-4}+10^{-6}k$
that satisfies $f(s,k) \in (0,1)$ and is strictly decreasing with
respect to $s$.

\subsection{Resilient Consensus}
We first consider the resilient consensus problem with bounded
noise. In the simulation setting, there are totally $n=10$ agents,
each of which has an arbitrarily selected initial state between $0$
and $1000$. The noise upper bound $\omega=10$. The random variable
for PFN has a continuous uniform distribution of the probability
density function
\begin{align*}
f_{\rm Random}(s)= \left\{
\begin{array}{ll} 1/1000, & s\in [0, 1000]\\
0, & {\rm otherwise}
\end{array}
\right. .
\end{align*}

\subsubsection{Fixed Topology Case}
\begin{figure}[!t]
\centering
\includegraphics[height = 0.54 \linewidth, width= 0.7 \linewidth]{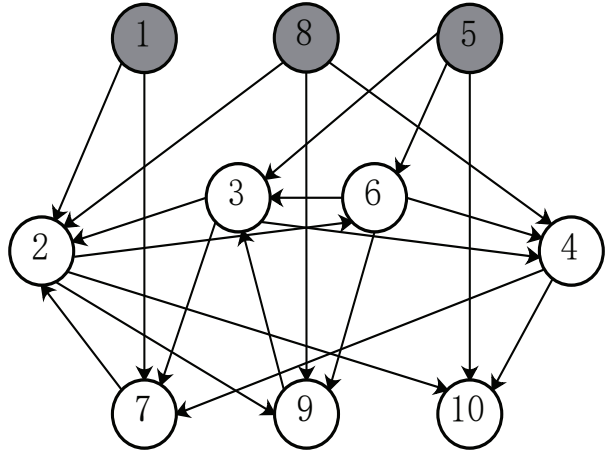}
\caption{A fixed communication topology where $1$, $5$ and $8$ are
faulty nodes, and the rest normal nodes are rooted.
 \label{Fig:topology1}}
\end{figure}

\begin{figure}[!t]
\centering
\psfrag{CostFunction}[][][0.7]{$V(\mathbf{x}(k))$}
\psfrag{Iteration}[][][0.7]{Iteration}
\includegraphics[width= 1.2 \linewidth]{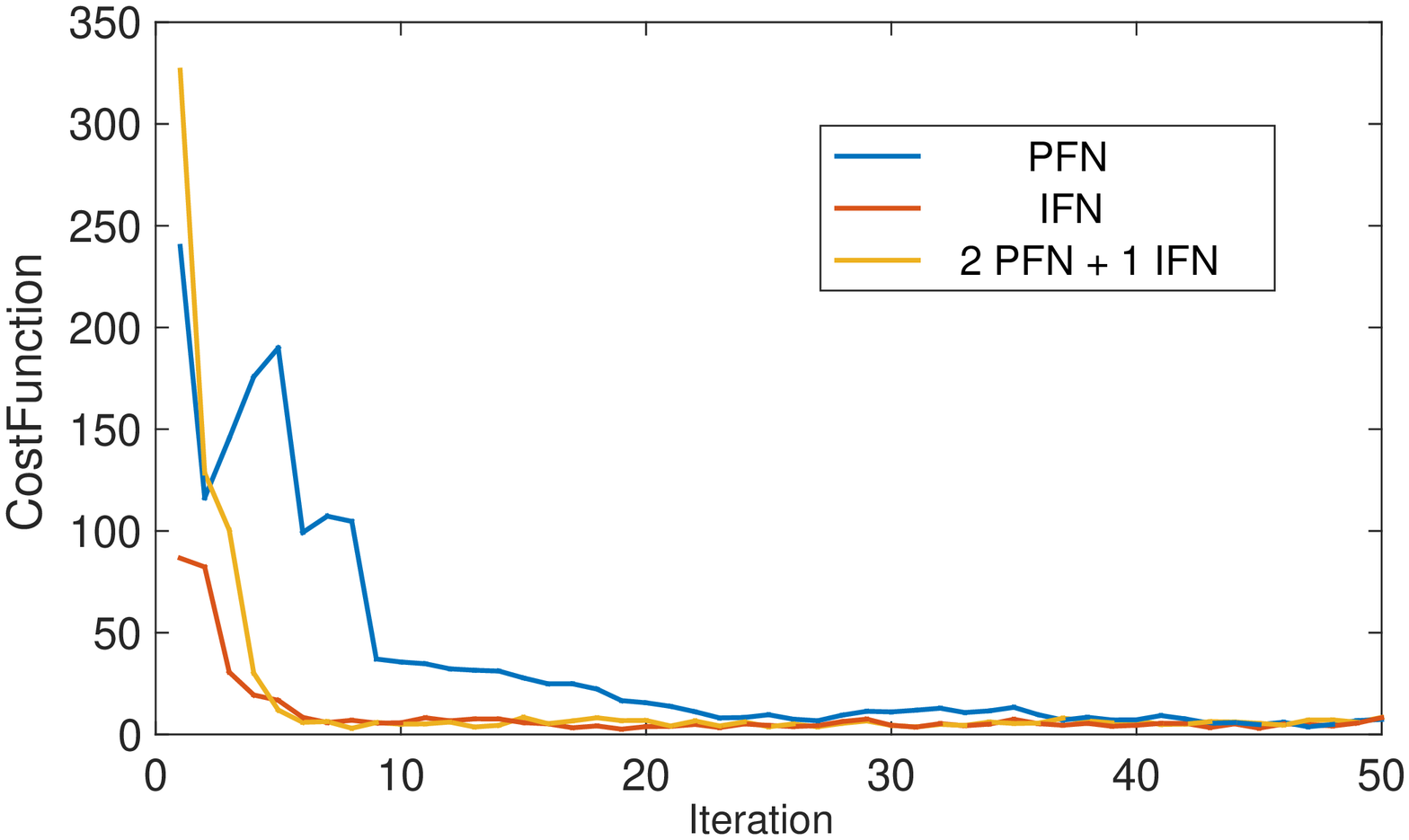}
\caption{Reaching resilient consensus for normal nodes using WLA-F.
 \label{Fig:sim1}}
\end{figure}

A fixed topology is given in Fig.~\ref{Fig:topology1} in which $1$,
$5$ and $8$ are faulty nodes, and all the rest are rooted normal
nodes. It is observed that for normal nodes $2$ and $6$, half of
their neighbors are faulty nodes such that the existing MSR
algorithms are unavailable. We update the adjacent weights by WLA-F
in the cases of PFN, IFN and the mixture of two, respectively, and
use
\begin{align*}
V(\mathbf{x}(k)) = \sqrt { \frac{1}{|V^{\rm n}| (|V^{\rm
n}|-1)}\sum_{i \neq j \in  V^{\rm n}} (x_i(k) - x_{j}(k))^2 }
\end{align*}
as a metric to assess the system convergence. Here, $|V^{\rm n}|$ is
the cardinality of $V^{\rm n}$.  The result is presented in
Fig.~\ref{Fig:sim1} showing resilient consensus in all three cases,
in which each IFN behaves normally with probability $p=0.8$ and
produces a random value otherwise.


\begin{figure}[!t]
\centering
\psfrag{Iteration}[][][0.7]{Average Convergence Number}
\psfrag{rate}[][][0.7]{Fault Probability}
\includegraphics[ width= 1.2 \linewidth]{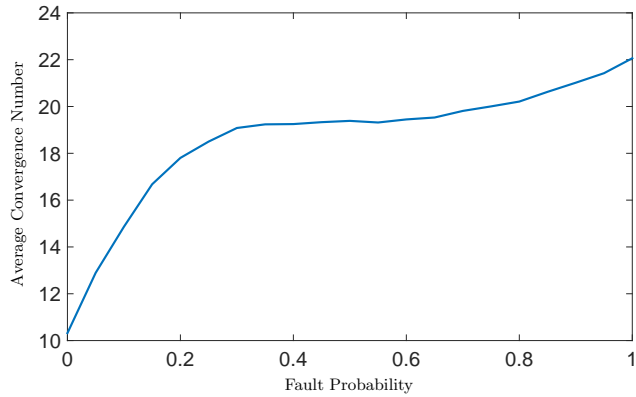}
\caption{The average convergence count varies with fault
probability.
 \label{Fig:sim11}}
\end{figure}

Next, we focus on the case of $3$ IFNs. In the above simulation, we
define that the fault probability of each IFN is $1-p= 0.2$. By
increasing this fault probability from $0$ to $1$ for each IFN, with
each round repeated $5000$ times, the average convergence count is
described in Fig.~\ref{Fig:sim11} where the metric $V$ smaller than
$5$ (half noise upper bound) is defined as convergence achievement.
From the simulation results, the convergence count grows with
increasing fault probability. More specifically, it  grows fast with
increasing fault probability at the range $[0, 0.2]$, and keeps flat
during $[0.2,0.8]$, and then continues to grow. This phenomenon is
due to the reason that small probability fault (occasional faulty
behavior) behind normal actions is more easily to be identified, and
adjacent weight from faulty nodes continues to drop with slightly
slower speed when fault probability increases.

\subsubsection{Stochastic Topology Case}

\begin{figure}[!t]
\begin{center}
\psfrag{CostFunction}[][][0.7]{$V(\mathbf{x}(k))$}
\psfrag{Iteration}[][][0.7]{Iteration}
\includegraphics[width= 1.2 \linewidth]{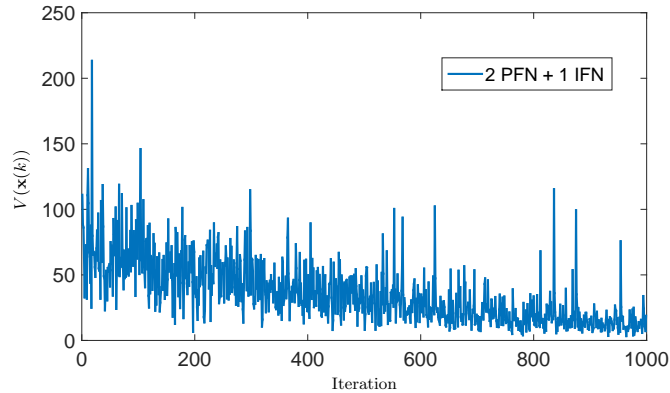}
\caption{Reaching resilient consensus for normal nodes using WLA-S.
 \label{Fig:sim2}}
\end{center}
\end{figure}

In the following, we assume that all nodes connect with each other
in probability $0.5$ at each step. Thus, the topology is rooted in
probability one. We define the parameter $\gamma=0.8$. We update the
adjacent weights by WLA-S, then the result presented in
Fig.~\ref{Fig:sim2} again validates our algorithm. The adjacent
weight after $1000$ iterations is given in Table I showing that the
adjacent weight from faulty nodes is quite small compared to the
ones from normal nodes. It should be noted that the column sum may
be greater than $1$, since for each node $i$, the weight sum of its
connected neighbors' at last iteration equals $0.8$, and all the
other weight values are kept since last connected time. Again, it
validates that IFNs can be more easily identified than PFNs.

\begin{table}
\centering \caption{Adjacent Weight}
\begin{tabular}{|c|c|c|c|c|c|c|c|}
\hline
&2 &  3 & 4 & 6 & 7 & 9 & 10 \\
\hline 1 & $\mathbf{0.019}$ & $\mathbf{0.027}$ & $\mathbf{0.021}$ &
$\mathbf{0.040}$ & $\mathbf{0.027}$ & $\mathbf{0.020}$ &
$\mathbf{0.027}$ \\[1pt]
2 & $0$ & $0.256$ & $0.190$ & $0.325$ & $0.229$ & $0.191$ & $0.261$ \\[1pt]
3 & $0.177$ & $0$ & $0.167$ & $0.393$ & $0.261$ & $0.197$ & $0.249$ \\[1pt]
4 & $0.152$ & $0.256$ & $0$ & $0.373$ & $0.249$ & $0.189$ & $0.258$ \\[1pt]
5 & $\mathbf{0.021}$ & $\mathbf{0.031}$ & $\mathbf{0.026}$ & $\mathbf{0.041}$ & $\mathbf{0.029}$ & $\mathbf{0.026}$ & $\mathbf{0.030}$ \\[1pt]
6 & $0.142$ & $0.269$ & $0.191$ & $0$ & $0.241$ & $0.186$ & $0.261$ \\[1pt]
7 & $0.163$ & $0.276$ & $0.200$ & $0.373$ & $0$ & $0.206$ & $0.267$ \\[1pt]
8 & $\mathbf{0.007}$ & $\mathbf{0.009}$ & $\mathbf{0.008}$ & $\mathbf{0.011}$ & $\mathbf{0.009}$ & $\mathbf{0.009}$ & $\mathbf{0.012}$ \\[1pt]
9 & $0.153$ & $0.257$ & $0.199$ & $0.333$ & $0.199$ & $0$ & $0.239$ \\[1pt]
10 & $0.169$ & $0.255$ & $0.190$ & $0.375$ & $0.225$ & $0.197$ & $0$ \\
\hline
\end{tabular}
\end{table}

In addition, we consider a large-scale network composed of $1000$
nodes, and the fault probability of IFN is $0.2$. By increasing the
IFN number from $1$ to $500$ with each round repeated $1000$ times,
the simulation results manifest that the normal nodes achieve
resilient consensus by a large majority. This is due to the the
reason that once abnormal behavior from IFN is recognized (large
relative state), the corresponding reward  and credibility  are
decreased and thus arduous to be promoted. This result indicates
that our algorithms work well in large-scale networks.

\subsection{Clock Synchronization}
\begin{figure}[!t]
\begin{center}
\includegraphics[ width= 0.7 \linewidth]{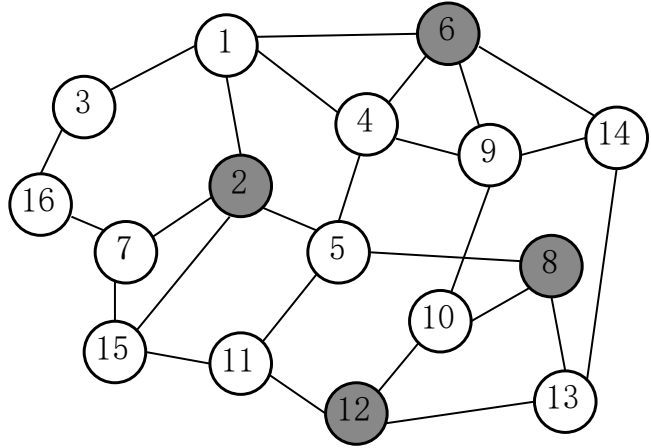}
\caption{A fixed communication topology where $2$, $6$, $8$ and $12$
are faulty nodes, and the rest normal nodes are rooted.
 \label{Fig:topology2}}
\end{center}
\end{figure}

\begin{figure}[!t]
\begin{center}
\psfrag{tau}[][][0.8]{$\tau$} \psfrag{Iteration}[][][0.7]{Iteration}
\includegraphics[width= 1.2 \linewidth]{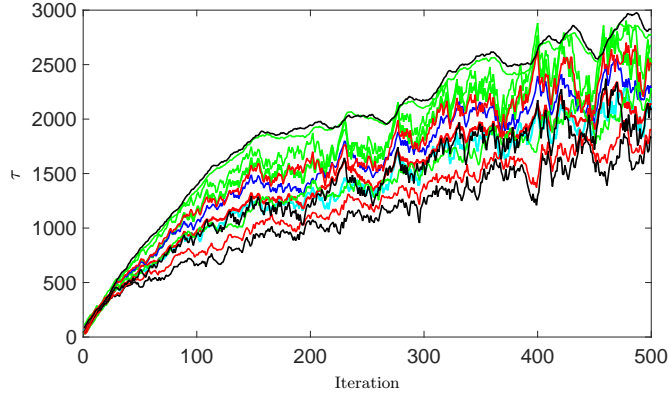}
\caption{Divergent phenomenon of a group of normal nodes caused by
faulty nodes without using WLA.
 \label{Fig:appunclock}}
\end{center}
\end{figure}

\begin{figure}[!t]
\begin{center}
\psfrag{tau}[][][0.8]{$\tau$} \psfrag{Iteration}[][][0.7]{Iteration}
\includegraphics[width= 1.2 \linewidth]{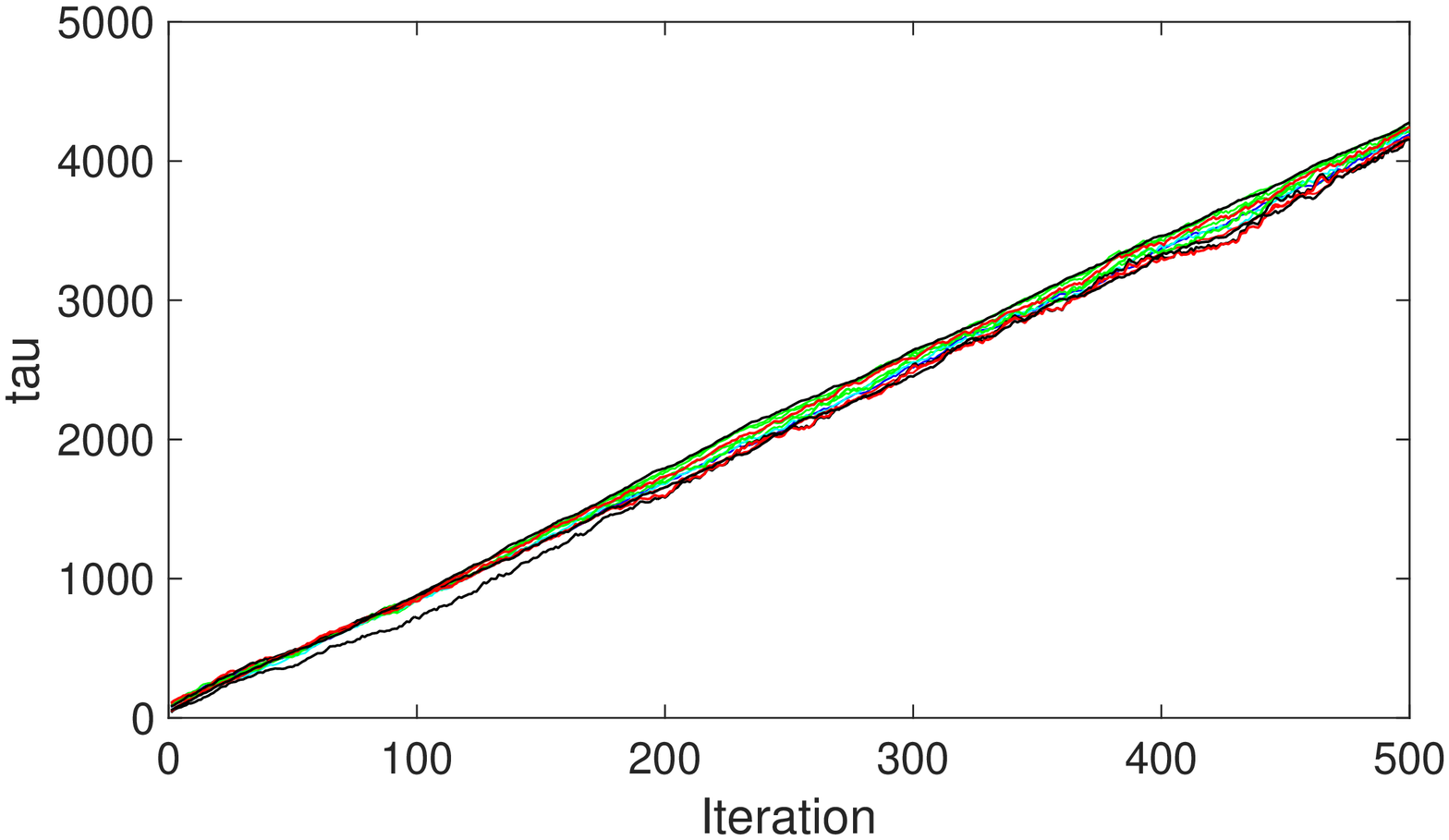}
\caption{Reaching resilient clock synchronization for each normal
node using WLA-F with fixed communication connection.
 \label{Fig:app}}
\end{center}
\end{figure}

In this subsection, we apply our approach in a clock synchronization
problem. We follow the example in \cite{FTCS17}, in which a sensor
network consists of $16$ nodes including $2$ PFNs and $2$ IFNs. The
undirected communication topology is shown in
Fig.~\ref{Fig:topology2} such that the MSR algorithms are not
available. The clock skew $\alpha^*_i$ and offset $\beta^*_i$ are
randomly initialized within the intervals $[0.7, 1.3]$ and $[0,
100]$, respectively. The noise upper bound is restrict as $5$. To
simplicity of experiment, the updating time instant is set $1$. The
initial parameter $\alpha_i(0)=1$ and $\beta_i(0)=0.1$ for all $i$.
The probability density functions $f_{\rm Random}$ for $\alpha$ and
$\beta$ in PFN follow a continuous uniform distribution over the
ranges $[0,5]$ and $[0,50]$, respectively. The logical clock
demonstration without and with WLA are shown in
Fig.~\ref{Fig:appunclock} and Fig.~\ref{Fig:app}, respectively,
indicating resilient clock synchronization can be achieved by our
approach, i.e., the logical clock difference between normal nodes is
bounded.

\section{Conclusions}\label{sec:con}
In this paper, we present WLA to solve the resilient multi-agent
consensus problem in the presence of noise and faulty nodes. Through
defining a reward function by relative neighbor nodes' states,
adjusting the corresponding credibility and adjacent weights, the
normal nodes finally achieve resilient consensus with reducing the
weights from faulty nodes to sufficiently small values. In our
study, two types of faulty behavior models and two types of network
topology are considered. Moreover, the proposed approach is applied
in a clock synchronization problem such that hacked or
malfunctioning nodes can be gradually ignored. By our algorithms,
the network condition is greatly relaxed and only one-step
information is required. In the future, we will consider the
cooperation among the faulty nodes and more applications in, e.g.,
beamforming technology and social networks.

\bibliographystyle{unsrt}
\bibliography{consensus}
\end{document}